\documentclass[letterpaper, 10 pt, conference]{ieeeconf}  

\IEEEoverridecommandlockouts                              

\usepackage{graphicx} 
\usepackage{subfigure}
\usepackage{amsmath} 
\usepackage{amssymb}  
\usepackage{url}

\title{\LARGE \bf
Adaptive motor control and learning in a spiking neural network realised on a mixed-signal neuromorphic processor
}

\author{ Sebastian Glatz$^{1}$, Julien Martel$^{2}$, Raphaela Kreiser$^{2}$, Ning Qiao$^{2}$ and Yulia Sandamirskaya$^{2}$
\thanks{*This work was supported by the SNSF grant PZOOP2\_168183, Ambizione, and a Fellowship of  the Neuroscience Center Zurich (ZNZ)}
\thanks{$^{1}$Sebastian Glatz is with Department of Electrical Engineering and Information Technology of ETH Zurich, Switzerland,
        {\tt\small sglatz@student.ethz.ch}}%
\thanks{$^{2}$Julien Martel, Raphaela Kreiser, Ning Qiao, and Yulia Sandamirskaya are  with the Institute of Neuroinformatics, University of Zurich and ETH Zurich, Switzerland,
        {\tt\small yulia.sandamirskaya@ini.uzh.ch}
}}

\begin{document}

\maketitle
\thispagestyle{empty}
\pagestyle{empty}

\begin{abstract}

 Neuromorphic computing is a new paradigm for design of  both the computing hardware and algorithms inspired by biological neural networks. The event-based nature and the inherent parallelism make neuromorphic computing a  promising paradigm for building efficient neural network based architectures for control of fast and agile robots. In this paper, we present a spiking neural network architecture that uses sensory feedback to control rotational velocity of a robotic vehicle. When the velocity reaches the target value, the mapping from the target velocity of the vehicle to the correct motor command, both represented in the spiking neural network on the neuromorphic device, is autonomously stored on the  device using on-chip plastic synaptic weights. We validate  the controller using a wheel motor of a miniature mobile vehicle and inertia measurement unit as the sensory feedback and demonstrate online learning of a simple ``inverse model'' in a two-layer spiking neural network on the neuromorphic chip.  The prototype neuromorphic device that features 256 spiking neurons allows us to realise a simple proof of concept architecture for the purely neuromorphic motor control and learning. The architecture can be easily scaled-up if a larger neuromorphic device is available.  

\end{abstract}

\section{INTRODUCTION}

  The ability to learn how to control a motor in order to achieve a goal state is an important component of any robotic system that should function under real-world conditions. When moving towards robots capable to adapt to changing environmental conditions and new tasks, robust and fast perception and learning become critical components. While motor controllers are typically implemented in small digital processors, the state of the art perception and learning increasingly use artificial neural networks (ANNs) \cite{LeCun2015} that require substantial computing power. Indeed, ANNs often require Graphical Processing Units (GPUs) to run in real time.  The power-latency trade-off is one of the challenges that hinder application of neural networks in online learning and adaptation of the sensory-motor mappings, required for adaptive motor control. Another limitation is the training procedure that requires a large amount of well-prepared data to drive the slow learning process using error back-propagation.
     
Neuromorphic processors are a new type of computing hardware that enables a radically low-power and compact realisation of spiking neural networks \cite{IndiveriEtAl2011,Indiveri2009,Benjamin2014}, which opens way to their application in power-critical domains, e.g., for fast and agile robots~\cite{FalangaEtAl2017,Mueggler2017}.  Spike-based learning and computing techniques are being developed that have been shown to demonstrate at least the same performance as ANNs \cite{dada,Neftci2017}. 

In this work, a spiking neural network was realised on a mixed signal neuromorphic chip. The neural network implements  feedback PI-control to converge on the correct motor command that achieves the goal speed, using sensory feedback. When the feedback controller settles on the desired task velocity, the mapping between the goal state and the motor control signal, both represented in a neural network on chip, is stored using on-chip synaptic plasticity. Since the integration of the spike-based neuromorphic and conventional digital hardware that typically controls motor systems, diminishes the advantages of low-power analog computing and event-based parallel processing, we  implement the  full system --both the feedback controller and  learning of the sensorimotor mapping-- in a neuromorphic device. This led us to a first neuromorphic realisation of a PI-controller. While several perception and  robotic architectures were introduced recently using mixed-signal neuromorphic devices \cite{BinasEtAl2016,Corradi2015,KreiserEtAl2017,Milde2017}, their applicability for closed-loop motor control has not been shown yet. The analog, subthreshold circuits are known to suffer from mismatch of computing elements and limited number of parameters to configure spiking neural networks \cite{Neftci2011}, making their use for motor control -- which requires fast and precise feedback -- challenging. Here, we found a solution to this problem.
     
 To achieve this, we developed a neuronal architecture that allows to control the speed of the robot by computing both differences and sums of sensed and desired values reliably in a spiking neural network. We demonstrate functionality and robustness of this architecture in a closed loop on a miniature robotic vehicle controlled by  the neuromorphic device. Most of computation is done in the spiking on-chip network here, only communication with the robot is accomplished through software. Doing this, we show that digital computing can potentially be taken out of the loop, leading to ultra low-power motor controllers that are easy to integrate with spiking neural networks for perception and higher-level cognitive processing \cite{Sandamirskaya2014}. Furthermore, we show how learning different movement commands can be realised using plastic -- adaptive -- synapses on the neuromorphic device in an online learning process that co-occurs with behavior.

We demonstrate the proof of concept on a small prototype neuromorphic device that features only 256 neurons and two arrays of 16K synapses realised with analogue electronics \cite{Qiao2015}. We consider this demonstration, however, an important step towards fully neuromorphic robotic controllers. Such controllers and learning systems will be orders of magnitude more power efficient, allowing us to scale up even complex cognitive neural network architectures. Larger neuromorphic devices became available recently \cite{Davies2018,Merolla2014} and scaling up our model  is straightforward due to the locality of connectivity leading to inherent parallellism of the computing architecture. The controller can easily be extended to control of effectors with several degrees of freedom, based on methods of spike-based motor control developed recently  \cite{Perez-PenaEtAl2014b,Perez-PenaEtAl2014,DeWolfEtAl2017,MenonEtAl2014}.


\section{METHODS}

\label{sec:methods}

\subsection{Overall setup and the motor control task}

\begin{figure}
  \centering
  \includegraphics[width=0.5\textwidth]{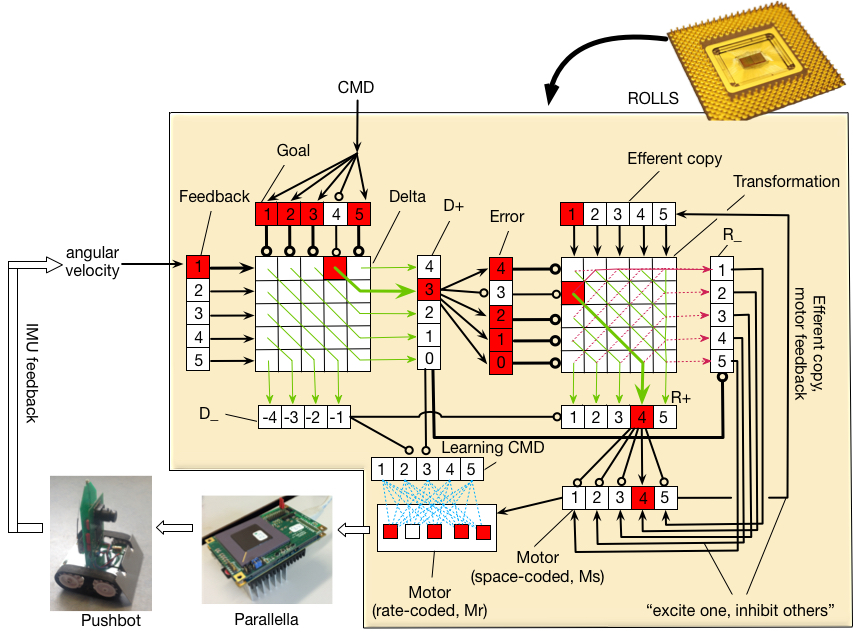} 
  \caption{Schematics of the spiking neural network for control of a single motor. Squares show silicon neurons, red squares are neurons that would be typically active in a control task. Arrows show excitatory connections, lines ending with circles -- inhibitory ones. Roles of different neural populations are explained in the main text.}
  \label{fig:controller}
\end{figure}

In our experiments, we used a setup shown in Fig.~\ref{fig:controller}, that consists of a neuromorphic device ROLLS (Section~\ref{sec:rolls}), mounted back-to-back via expansion connectors to a miniature computer Parallella \cite{Olofsson2015}. The Parallella is  used to configure the device, monitor its activity, and direct events (spikes) between ROLLS and the robot. The miniature robotic vehicle Pushbot\footnote{\url{https://inilabs.com/products/pushbot/}} consists of a pair of propulsion chassis and an embedded Dynamic Vision Sensor (eDVS, \cite{MullerEtAl2011}) with an artificial retina DVS that embeds a 6-axis IMU and a microcontroller managing wireless communication using a serial protocol. While we have used WiFi connection, the Parallella board, and the microcontroller of the eDVS in our prototype system to set the instantaneous speed of the motor, a more integrated solution with ROLLS directly driving the motor could use recently introduced models for open-loop control with spike-based pulse-frequency modulation \cite{Perez-PenaEtAl2014b,Perez-PenaEtAl2014}.

Our experiments consisted in selecting a number of rotation velocities that the robot should adopt and letting the neuromorphic controller set these velocities, based on a feedback from IMU. In a second set of experiments, we additionally learned the control signals on chip.

\subsection{Neuromorphic device ROLLS}
\label{sec:rolls}
 Two types of neuromorphic platforms exist: (1) purely digital systems, optimised for running spiking neural networks simulations   \cite{FurberEtAl2012,Davies2018,Merolla2014} and (2) mixed-signal systems that use analog circuits to emulate in silicon the  computational neuroscience models \cite{Benjamin2014,SchemmelEtAl2010,Indiveri2009}. Some of the latter devices were optimised in their small form factor and ultra low power consumption for robotic, i.e. real-time and embedded, applications \cite{Qiao2015,MoradiEtAl2017,Milde2017}. 
 
 
The neuromorphic device ROLLS (Reconfigurable OnLine Learning Spiking) \cite{Qiao2015} is such a mixed signal Very Large Scale Integration (VLSI) neuromorphic device  featuring 256 spiking neurons realised with analog electronic circuits \cite{IndiveriEtAl2011} that emulate adaptive leaky integrate-and-fire neuron model. The neurons on chip  can be connected by a set of all-to-all (16K) synapses, also realised with analog circuits, low-pass filtering incoming (digital) spikes. Another array of 16K plastic (adaptive) synapses can be used to realise on-chip learning rule. The learning rule is a version of STDP (spike-timing dependent plasticity) rule, which is particularly well-suited to hardware realisation and can be used to implement simple Hebbian learning on population level (``fire together -- wire together")~\cite{Fusi2000, Mitra2009, Indiveri2006}. ROLLS was produced using 180nm CMOS technology, occupies an area of of 51.4mm$^2$, and consumes $<$4mW of power when all neurons fire at a typical rate of 30Hz. 

Communication with the device is implemented using Address Event Representation (AER) \cite{Boahen99}. Thus, each spike is sent both within the chip and to other devices as a digital package -- an ``event'' -- that contains the neuron's index (0-255) and a time-stamp.  ROLLS also receives AER packages as inputs, with flags determining whether an in-coming event is a stimulus, or a bias-setting used for configuring the chip. 

``Programming" the ROLLS chip amounts to (1) assigning different roles to different neurons by directing sensory input to some of them and using activity of other ones to drive the robot's motors; (2) specifying non-plastic connections between neurons by activating required synapses and setting their parameters (one of 4 positive or 4 negative weights and other parameters such as time-constants and thresholds of the low-pass filters); and (3) activating, if needed, some of the plastic synapses for on-chip learning. 

\subsubsection{Coping with device missmatch}

Since ROLLS uses analog, subthreshold circuits, the computing elements are subject to device mismatch, which is one the main challenges in working with mixed signal (and thus particularly low power and compact) neuromorphic devices \cite{Neftci2011}. Consequently,  with identical parameter settings, the silicon neurons show different behaviour. 

\begin{figure}
  \centering
  \includegraphics[width=0.5\textwidth]{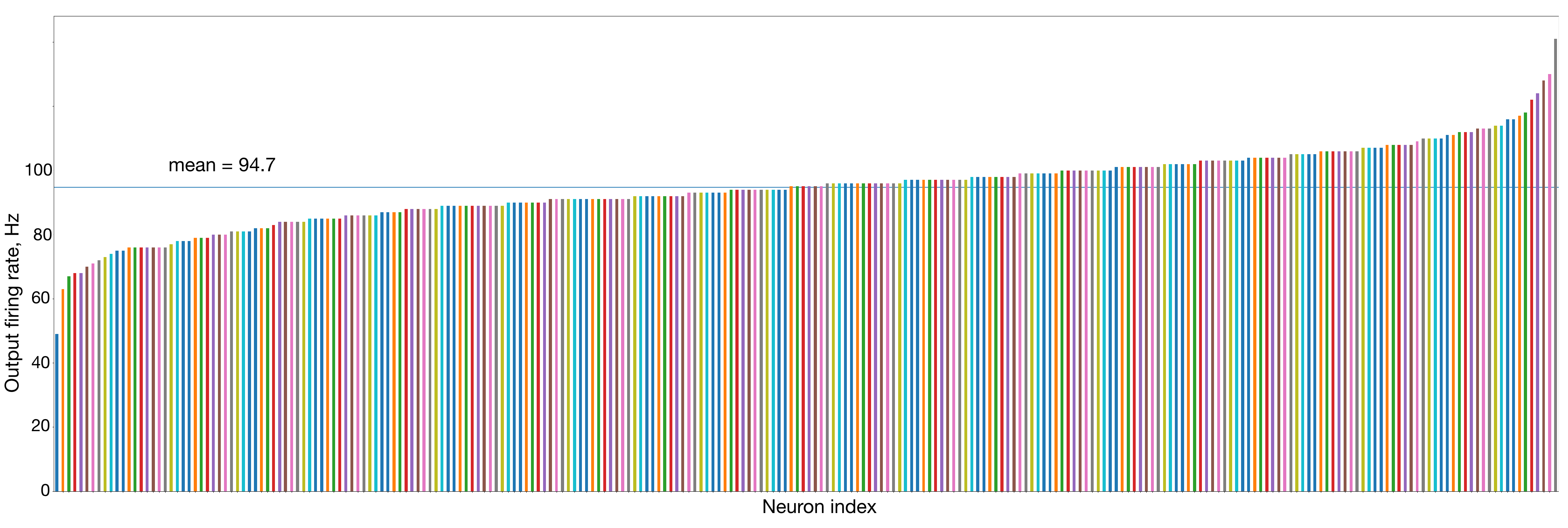} 
  \caption{Firing rate of spiking neurons on the ROLLS chip when each neuron is stimulated with 200Hz, ordered.}
  \label{fig:neurons_sweep}
\end{figure}

Fig.~\ref{fig:neurons_sweep} shows distribution of firing rates of 256 neurons on the ROLLS chip when they are stimulated externally with 200Hz and have identical parameters.  One can see that output firing rates, ordered in the plot, cover a wide range of values (mean 94.7Hz, standard deviation 11.9Hz), which would make behavior of spiking neural networks extremely noisy if one assumes homogeneous behavior of neurons. Here, we developed a routine to select similar neurons for the neural populations in our controller architecture based on the measured distribution. Since behavior of neurons over time is rather stable, this approach leads to robust control architecture, reducing the effect of mismatch of the neuromorphic device.

\subsection{The neuronal controller architecture}

In order to control the speed of the robot, we used a simple proportional controller, realised with a spiking neural network, shown in Fig.~\ref{fig:controller}.

In the network, we set up different neural populations to represent: the (G)oal -- the task-level command, i.e. the desired rotational speed of the robot; (F)eedback -- the current rotation speed, as signaled by the IMU; (D)elta -- a 2D neural array that computes the difference between (G) and (F); Read-out populations that represent a positive (D$_+$) or a negative (D$_-$) difference between the command and the IMU-measurement; (Ms) -- motor output represented with ``space-code'', in which each active neuron represents a different motor command; (T)ransformation -- a 2D neural array that maps delta values onto motor command values; (Mr) -- motor output in a ``rate-code'' that represents the motor command by the firing rate that directly drives the motor. 

The final motor command is proportional to the instanteneous firing rate of the Mr population, i.e. the spike-count in 100ms time window.  A more direct spike-based code can be used to control the motors in a more embedded neuromorphic realisation, as has been explored recently \cite{Perez-PenaEtAl2014b,Perez-PenaEtAl2014,DonatiEtAl2018}. Our architecture adds space-coded representations on top of the rate-coded one:  in the the Ms population, each neuron represents a different motor command.  This enables online learning of a mapping between the task and motor command space using simple Hebbian learning rule of the on-chip plastic synapses.

Thick arrows in Fig.~\ref{fig:controller} show a typical activation flow in the controller from the externally set task command (Goal) to activation of the motor rate-coded population (Mr). Red squares show neurons that would be active in a typical control task: 

(1) one of the Goal-neurons is inhibited by an external signal, setting the desired rotational velocity of the robot. We used a 
\emph{double-inhibition connectivity pattern} in several places in the architecture, which leads to more robust behavior of analog spiking neurons than an alternative excitatory connectivity. The latter would require precise summation of two inputs in the 2D array (e.g., the Delta or Transformation arrays here), which, in our experiments, was shown to be hard to tune, because of the narrow parameter range, for which only the neuron, in which the two inputs ``cross'', fires. Interestingly, such double-inhibition connectivity pattern is pervasive in the brain, in particular in cortico-basal ganglia circuits, involved in action control \cite{Yin2006,Aldridge2003}. 

(2) The inhibited Goal-neuron releases inhibition of the respective column of the ``Delta'' neuronal array. A row in this array is selected by an excitatory connection from the IMU-population, signalling the current speed of the robot, leading to activation of a single neuron in the 2D array. 

(3) Read-out from the Delta-array is wired-up in such a way that two 1D output arrays (D$_+$ and D$_-$) represent the absolute difference between the positions of an active (or inhibited) neuron in the input arrays (Feedback and Goal, respectively) in positive or negative direction, respectively. Thus, the 2D delta-array is a neuronal ``operator'' that realises a particular \emph{relation} between its inputs and outputs (``difference'' here). Such neuronal operators were introduced  in  neural networks simulating related brain function \cite{Pouget2000} and were used in neuronal cognitive architectures for coordinate transformations \cite{SandamirskayaConradt2013_ICANN} and for representing spatial relations \cite{RichterEtAl2014}. Recently it has been shown how such operators, in principle, can be learned \cite{Diehl2016}. This opens a possibility for development of fully adaptive neuromorphic  controllers. 

(4) Activity in the D$_+$ and D$_-$ neural arrays is mutually exclusive and they inhibit the respectively opposite output array of the second ``operator'' 2D array -- the transformation array. The transformation array takes as input the computed difference between the set and the actual velocities and the current motor command, sent to the motors (such feedback of the current motor command is called ``efferent copy'' in computational neuroscience literature, the term we also use here). This time, the wiring of the 2D array is set to sum the activity of its neurons along left- and right-diagonals, leading to computing either a sum or a difference of its inputs, depending on whether the D$_+$ or the $D_-$ population is active.

(5) Outputs (``results'') of the Transformation array -- $R_+$ and $R_-$ -- drive a space-coded motor population (Ms), which represents the motor command in activity of \emph{one} of its neurons (each neuron representing a different motor command). A winner-take-all connectivity pattern (realised by an auxiliary inhibitory neural array, not shown in Fig.~\ref{fig:controller} to avoid clutter) ensures that only one Ms neuron can be active at a time. 

(6) Finally, each motor neuron is connected to a different number of rate-coding motor neurons (Mr), depending on the represented motor command. Thus, the first space-coded motor neuron is connected to one rate-coded neuron, whereas the fifth space-coded neuron is connected to 5 rate-coded neurons that can be selected arbitrarily from the rate-coded motor array. The more neurons in the Mr array fire, the higher the overall firing rate of this neural population. This firing rate sets the instantaneous speed of the motor (see Section~\ref{sec:software}). 

In Section~\ref{sec:res}, we show how this controller is able to set the desired rotation speed of the robot, using feedback from IMU. Feedback from event-based motion detection based on eDVS output can be used as well and will be reported shortly.

\subsection{Interfaces computed in software}
\label{sec:software}

We have used a thin software layer, running on the Parallella board, to establish interfaces between the robot and the spiking neurons on the ROLLS chip. Thus, the rotational velocity,  measured at 200Hz using the IMU,  was binned in one of 5 ranges. Depending on the interval in which the measured velocity fell, one of 5 IMU (Feedback) neurons was stimulated with a rate of 800Hz. On the other side, the activity of the Motor rate-coded array was calculated by counting spikes in 100ms time windows. The speed command sent to the robot was set proportionally to this spike-rate. This software solution was used in our prototype system that didn't yet have a direct connection to the robot. When such connection is established, direct spike-based control will be possible \cite{Perez-PenaEtAl2014b,Perez-PenaEtAl2014,DonatiEtAl2018}. The CMD signal (the goal velocity) was set by stimulating inhibitory synapses of one of the Goal neurons.     
    
\subsection{Learning feedforward control}

Additionally to the feedback controller, we demonstrate here how a direct feedforward control signal can be learned on the neuromorphic device ROLLS, which features on-chip plasticity. Thus, we have added an array of plastic synapses between the learning command neural population and the rate-coded motor population (light blue dashed lines in Fig.~\ref{fig:controller}). We set up learning to be active when the Delta-value is zero (meaning that the feedback controller has converged): the learning command population is inhibited by all other Delta+/- neurons. During learning, the plastic synapses  potentiate  (increase) between the active learning command neuron and active Mr neurons. These synapses store a direct connection from the task-level command to the motor population, leading to faster activation of the correct motor behavior after learning. Both controllers can work in parallel, complimenting each other. Note that learning here is controlled by activity of the Delta=0 neuron, turning on-chip plasticity on and off during behavior. Such online learning in a neuromorphic device has been demonstrated for the first time here.


\section{RESULTS}

\label{sec:res}

\subsection{Adaptive control}

\begin{figure*}
  \subfigure[]{
  \includegraphics[width=0.49\textwidth]{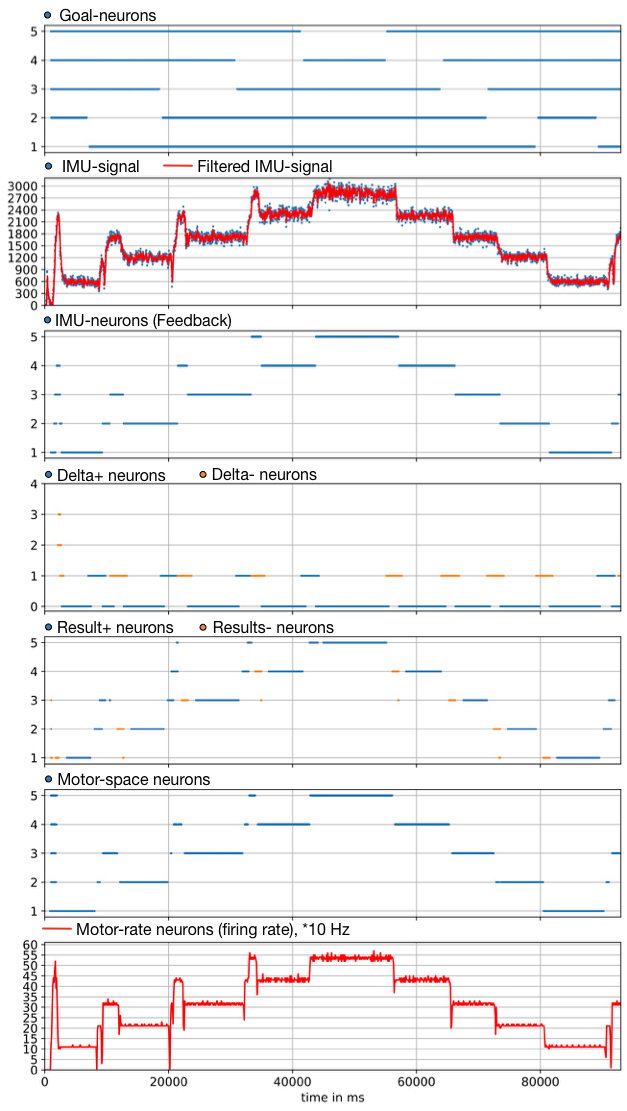}
  \label{fig:res1}} 
    \subfigure[]{
  \includegraphics[width=0.49\textwidth]{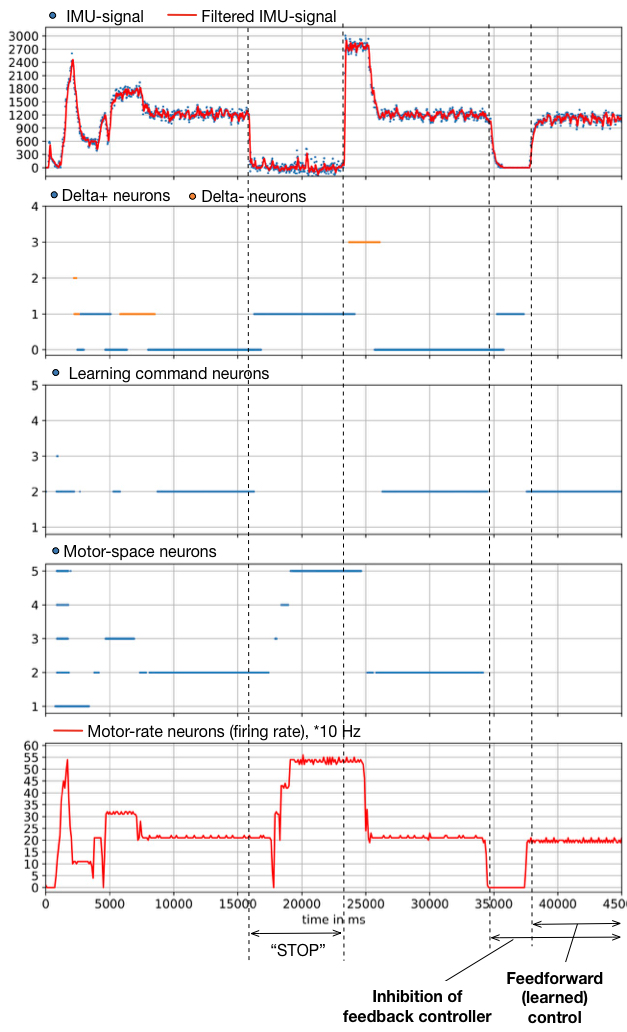} 
  \label{fig:res2}
  }
    \caption{(a) Time plots of activity of neuronal populations on the ROLLS chip and IMU measurements during a sequence of control tasks. (b) Activity of neural populations on the ROLLS chip and IMU measurements in a ``stop-and-go'' and learning experiment. See Section~\ref{sec:res} for details.}
\end{figure*}

Fig.~\ref{fig:res1} shows the output of the neuromorphic chip (spikes over time) in a sequence of control tasks. Here, we set control signals 1-5 to the chip in an increasing and decreasing order, with a step size of 1. Other step sizes  show similar results. Upper plot shows activity of 5 Goal neurons, one of which is inhibited and corresponds to the selected control signal (desired speed of the robot). The second plot shows IMU measurements of the robot's speed. One can note that the controller overshoots first on the rising steps, but then settles on the correct speed and keeps it. In the decreasing steps, the controller reaches the desired speed directly. The third plot shows activity of the Feedback neurons on the ROLLS chip, which are driven by the IMU signal and faithfully represent it. The forth and fifth plots show activity of the Delta+/- and Result+/- neurons, respectively. The controller has converged on a desired value when only Delta+ = 0 is active. The second to last plot shows activity of the space-coded motor neurons, driven by the Result+/- neurons, and the bottom plot shows firing rate of the rate-coded motor neurons, which drive the motor.

Overall, the figure shows how the spiking controller succeeds in setting the desired robot's speed by controlling its wheel motor. We have tested the controller for all step sizes and different target levels and have shown that it is in all cases capable to set the required speed. The time to reach the desired speed was on the order of 5s and 3s for the rising and decreasing speeds, respectively (measured on 10 up-and-down sweeps).  The high settling time comes from the simplicity of the controller, which only uses 5 values (neurons) for the control signal, the feedback, and the differences, which hinders the convergence, in particular when the measured IMU signal is on the boarder between the selected 5 bins. Moreover, our ``outer" feedback loop is rather slow, much slower than the ``inner" feedback loop (with the efferent copy of the motor command). This problem will be alleviated in an integrated neuromorphic solution.

\begin{figure}[t!]
  \centering
  \includegraphics[width=0.5\textwidth]{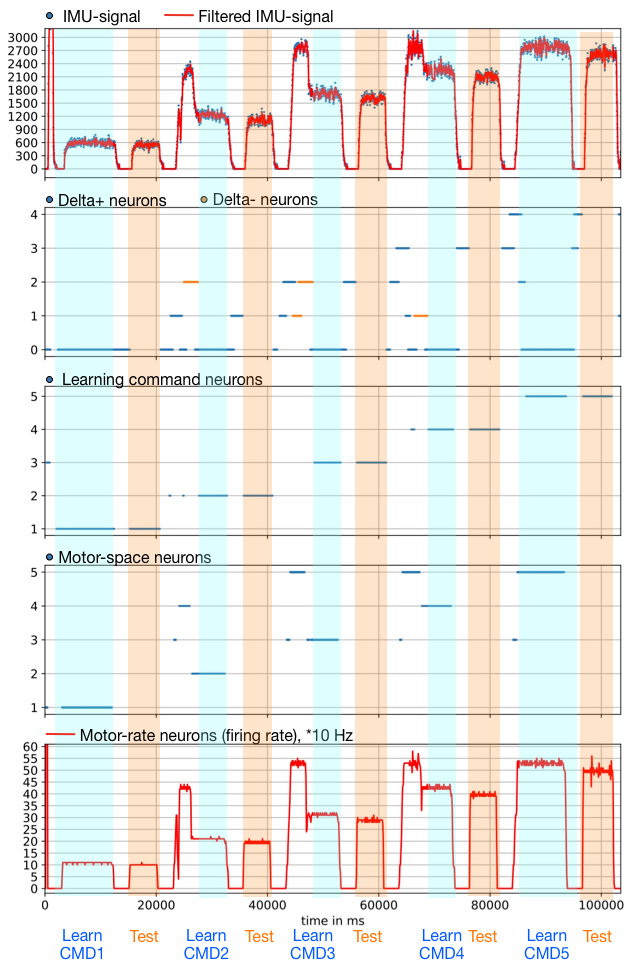} 
  \caption{Time plots of activity on the ROLLS chip during testing of the learned feedforward controller for all five commands.}
  \label{fig:res3}
\end{figure}

The fluctuations  in  speed of the robot when the controller has converged are on the order of 84 IMU measurement units (at the mean value of 1012, measured for 1 minute). These fluctuations stay unchanged in experiments without the controller (when setting a given speed and measuring IMU signal). Thus, they are largely determined by the precision of the IMU measurements and not the neuronal controller.


\subsection{Perturbations and learning movement goals}


Fig.~\ref{fig:res2} shows time-plot of activity of the controller in one of the experiments in which we demonstrate (1) workings of the controller if a perturbation is applied to the robot (stop-and-go task) and (2) functioning of the learned direct, feedforward control. 

Here, the top plot shows the IMU measurement during the experiment. We set the goal speed to 2 and wait for the controller to settle on this speed. At $\approx$16000ms, we hold back the robot, preventing it from turning (a ``STOP'' perturbation), releasing the robot after approx. 8 seconds. One can see how the feedback controller ramps-up the speed of the robot, trying to reach the set speed and failing to do so due to the disturbance. When the robot is released, the controller sets its speed back to the desired value. 

During this experiment, each time when the controller settles on the desired value (i.e. Delta neuron 0 is active) learning is activated, triggering learning in plastic synapses on the ROLLS chip. Plastic synapses between the active learning command and the active rate-coded motor neurons are strengthened then. In the example shown in Fig.~\ref{fig:res2}, just before the time mark 35000ms, we inhibit the feedback controller (note absent activity in the motor space-coded neurons, which drive the rate-coded motor neurons during the feedback control). The robot stops and when the learning command is activated (released from inhibition of the Delta neurons), the feedforward controller shows its effect. We can see that the robot's speed is immediately set to the desired value. This is achieved by activating the correct number of rate-coded neurons through plastic synapses from the learning command neurons (note that the only active neuron in the architecture after 37500ms is the learning command neuron, which drives the Mr neural population).

Fig.~\ref{fig:res3} shows learning of all 5 motor commands on the ROLLS chip. The light blue regions mark time periods when the feedback controller has converged to the set value and activates the learning command neurons. In these periods, plastic synapses between the learning command neurons and the active motor neurons are strengthened. During time periods marked orange in the figure, the feedback controller is inhibited to probe the feedforward connections. In these periods, the active learning command neurons (third plot) activate the Mr motor neurons (last plot) directly (no other neurons are active in the architecture). We show here that all commands can be learned and drive the motors directly (as can be seen in the IMU measurements of the robot, top plot).

\vspace{-0.3cm}
\section{CONCLUSION}
\label{sec:conclusion}
\vspace{-0.3cm}
In this paper, we have presented a new purely neuromorphic motor controller that sets desired speed (rotational velocity here) of the robot based on IMU-feedback in a closed loop. The controller is fully realised on a mixed-signal neuromorphic device that emulates spiking dynamics of neurons in subthreshold, analog circuits, consuming $<$4mW of power for the actual computation with spikes. While many neural-network based controllers rely on high-dimensional representations and large networks to do adaptive control \cite{DeWolfEtAl2017,MenonEtAl2014}, we follow a different approach here and design a ``minimal'' controller that uses only as many neurons as needed to solve the control task ($<$100 here). While the controller, presented here, was implemented on a small prototype device, following our approach, controllers for multiple motors and objectives could be easily realised on a larger neuromorphic system, such as SpiNNacker \cite{FurberEtAl2012}, TrueNorth \cite{Merolla2014}, or Loihi \cite{Davies2018}. Using adaptivity of the neural circuits, powerful adaptive controllers can be developed using this paradigm in the future, using well-known neural control methods, such as \cite{GuentherEtAl94,MenonEtAl2014}. Such neuronal controllers can be more naturally integrated with perceptual systems, for which neuronal networks, including their neuromorphic realisations, are increasingly and successfully used.

\section*{ACKNOWLEDGMENT}


\bibliographystyle{plain}
\bibliography{literature_mendeley,elisa}  

\begin{thebibliography}{10}

\bibitem{Aldridge2003}
J~Wayne Aldridge and Kent~C Berridge.
\newblock {Basal Ganglia Neural Coding of Natural Action Sequences}.
\newblock {\em The Basal Ganglia VI}, 54:65--73, 2003.

\bibitem{Benjamin2014}
Ben~Varkey Benjamin, Peiran Gao, Emmett McQuinn, Swadesh Choudhary, Anand~R.
  Chandrasekaran, Jean-Marie Bussat, Rodrigo Alvarez-Icaza, John~V. Arthur,
  Paul~a. Merolla, and Kwabena Boahen.
\newblock {Neurogrid: A Mixed-Analog-Digital Multichip System for Large-Scale
  Neural Simulations}.
\newblock {\em Proceedings of the IEEE}, 102(5):699--716, may 2014.

\bibitem{BinasEtAl2016}
Jonathan Binas, Daniel Neil, Giacomo Indiveri, Shih-Chii Liu, and Michael
  Pfeiffer.
\newblock {Precise deep neural network computation on imprecise low-power
  analog hardware}.
\newblock pages 1--21, 2016.

\bibitem{Boahen99}
Kwabena~A Boahen.
\newblock {Point-to-Point Connectivity Between Neuromorphic Chips using
  Address-Events}.
\newblock {\em Ieee Transactions on Circuits {\&} Systems}, 47(5):416--434,
  1999.

\bibitem{Corradi2015}
Federico Corradi and Giacomo Indiveri.
\newblock {Real-time classification of complex patterns using spike-based
  learning in neuromorphic VLSI}.
\newblock {\em IEEE Transactions on Biomedical Circuits and Systems},
  3(1):32--42, 2009.

\bibitem{Davies2018}
Mike Davies, Narayan Srinivasa, Tsung-han Lin, Gautham Chinya, Yongqiang Cao,
  Sri~Harsha Choday, Georgios Dimou, Prasad Joshi, Nabil Imam, Shweta Jain,
  Yuyun Liao, Chit-kwan Lin, Andrew Lines, Ruokun Liu, Deepak Mathaikutty,
  Steve Mccoy, Arnab Paul, Jonathan Tse, Guruguhanathan Venkataramanan, Yi-hsin
  Weng, Andreas Wild, Yoonseok Yang, and Hong Wang.
\newblock {Loihi: a Neuromorphic Manycore Processor with On-Chip Learning}.
\newblock {\em IEEE Micro}, 38(1):82--99, 2018.

\bibitem{DeWolfEtAl2017}
Travis DeWolf, Terrence~C. Stewart, Jean-Jacques Slotine, and Chris Eliasmith.
\newblock {A spiking neural model of adaptive arm control}.
\newblock {\em Proceedings of the Royal Society B: Biological Sciences},
  283(1843):20162134, 2016.

\bibitem{Diehl2016}
Peter~U. Diehl and Matthew Cook.
\newblock {Learning and Inferring Relations in Cortical Networks}.
\newblock {\em arXiv preprint arXiv:1608.08267}, 2016.

\bibitem{DonatiEtAl2018}
E.~Donati, F.~Perez-Pe\~na, C.~Bartolozzi, G.~Indiveri, and E.~Chicca.
\newblock Open-loop neuromorphic controller implemented on vlsi devices.
\newblock In {\em 7th IEEE RAS/EMBS International Conference on Biomedical
  Robotics and Biomechatronics (BioRob)}, 2018.

\bibitem{FalangaEtAl2017}
Davide Falanga, Elias Mueggler, Matthias Faessler, and Davide Scaramuzza.
\newblock {Aggressive quadrotor flight through narrow gaps with onboard sensing
  and computing using active vision}.
\newblock {\em Proceedings - IEEE International Conference on Robotics and
  Automation}, pages 5774--5781, 2017.

\bibitem{FurberEtAl2012}
Steve~B. Furber, David~R. Lester, Luis~A. Plana, Jim~D. Garside, Eustace
  Painkras, Steve Temple, and Andrew~D. Brown.
\newblock {Overview of the SpiNNaker System Architecture}.
\newblock {\em IEEE Transactions on Computers}, 62(12):2454--2467, 2012.

\bibitem{Fusi2000}
Stefano Fusi.
\newblock {Spike-Driven Synaptic Plasticity: Theory , Simulation , VLSI}.
\newblock 2258:2227--2258, 2000.

\bibitem{GuentherEtAl94}
F~Guenther, D~Bullock, D~Greve, and S~Grossberg.
\newblock {Neural representations for sensory-motor control III. Learning a
  body-centered representation of 3-D target position}.
\newblock {\em Journal of Cognitive Neuroscience}, 6:341--358, 1994.

\bibitem{Indiveri2006}
Giacomo Indiveri, Elisabetta Chicca, and Rodney Douglas.
\newblock {A VLSI Array of Low-Power Spiking Neurons and Bistable Synapses With
  Spike-Timing Dependent Plasticity}.
\newblock 17(1):211--221, 2006.

\bibitem{Indiveri2009}
Giacomo Indiveri, Elisabetta Chicca, and Rodney~J. Douglas.
\newblock {Artificial Cognitive Systems: From VLSI Networks of Spiking Neurons
  to Neuromorphic Cognition}.
\newblock {\em Cognitive Computation}, 1(2):119--127, 2009.

\bibitem{IndiveriEtAl2011}
Giacomo Indiveri, Bernab{\'{e}} Linares-Barranco, Tara~Julia Hamilton,
  Andr{\'{e}} van Schaik, Ralph Etienne-Cummings, Tobi Delbruck, Shih-Chii~C
  Liu, Piotr Dudek, Philipp H{\"{a}}fliger, Sylvie Renaud, Johannes Schemmel,
  Gert Cauwenberghs, John Arthur, Kai Hynna, Fopefolu Folowosele, Sylvain
  Saighi, Teresa Serrano-Gotarredona, Jayawan Wijekoon, Yingxue Wang, and
  Kwabena Boahen.
\newblock {Neuromorphic silicon neuron circuits.}
\newblock {\em Front Neurosci}, 5:73, 2011.

\bibitem{KreiserEtAl2017}
Raphaela Kreiser, Timoleon Moraitis, Yulia Sandamirskaya, and Giacomo Indiveri.
\newblock On-chip unsupervised learning in winner-take-all networks of spiking
  neurons.
\newblock In {\em Biological Circuits and Systems (BioCAS)}, 2017.

\bibitem{LeCun2015}
Yann LeCun, Yoshua Bengio, and Geoffrey Hinton.
\newblock {Deep learning}.
\newblock {\em Nature}, 521(7553):436--444, 2015.

\bibitem{dada}
Wolfgang Maass and Henry Markram.
\newblock On the computational power of circuits of spiking neurons.
\newblock {\em Journal of computer and system sciences}, 69(4):593--616, 2004.

\bibitem{MenonEtAl2014}
Samir Menon, Sam Fok, Alex Neckar, Oussama Khatib, and Kwabena Boahen.
\newblock {Controlling Articulated Robots in Task-Space with Spiking Silicon
  Neurons}.
\newblock {\em 5th IEEE RAS/EMBS International Conference on Biomedical
  Robotics and Biomechatronics}, pages 181--186, 2014.

\bibitem{Merolla2014}
Paul~A Merolla, John~V Arthur, Rodrigo Alvarez-icaza, Andrew~S Cassidy, Jun
  Sawada, Filipp Akopyan, Bryan~L Jackson, Nabil Imam, Chen Guo, Yutaka
  Nakamura, Bernard Brezzo, Ivan Vo, Steven~K Esser, Rathinakumar Appuswamy,
  Brian Taba, Arnon Amir, Myron~D Flickner, William~P Risk, Rajit Manohar, and
  Dharmendra~S Modha.
\newblock {Network and Interface}.
\newblock {\em Sciencemag.Org}, 345(7812):668--673, 2014.

\bibitem{Milde2017}
M.B. Milde, H.~Blum, A.~Dietm{\"{u}}ller, D.~Sumislawska, J.~Conradt,
  G.~Indiveri, and Y.~Sandamirskaya.
\newblock {Obstacle avoidance and target acquisition for robot navigation using
  a mixed signal analog/digital neuromorphic processing system}.
\newblock {\em Frontiers in Neurorobotics}, 11(JUL), 2017.

\bibitem{Mitra2009}
Srinjoy Mitra, Stefano Fusi, and Giacomo Indiveri.
\newblock {Real-time classification of complex patterns using spike-based
  learning in neuromorphic VLSI}.
\newblock {\em IEEE Transactions on Biomedical Circuits and Systems},
  3(1):32--42, 2009.

\bibitem{MoradiEtAl2017}
Saber Moradi, Ning Qiao, Fabio Stefanini, and Giacomo Indiveri.
\newblock {A scalable multi-core architecture with heterogeneous memory
  structures for Dynamic Neuromorphic Asynchronous Processors (DYNAPs)}.
\newblock (August), 2017.

\bibitem{Mueggler2017}
Elias Mueggler, Henri Rebecq, Guillermo Gallego, Tobi Delbruck, and Davide
  Scaramuzza.
\newblock {The event-camera dataset and simulator: Event-based data for pose
  estimation, visual odometry, and SLAM}.
\newblock {\em International Journal of Robotics Research}, 36(2):142--149,
  2017.

\bibitem{MullerEtAl2011}
Georg~R. M{\"{u}}ller and J{\"{o}}rg Conradt.
\newblock {A miniature low-power sensor system for real time 2D visual tracking
  of LED markers}.
\newblock {\em 2011 IEEE International Conference on Robotics and Biomimetics,
  ROBIO 2011}, pages 2429--2434, 2011.

\bibitem{Neftci2011}
Emre Neftci, Elisabetta Chicca, Giacomo Indiveri, and Rodney Douglas.
\newblock {A Systematic Method for Configuring VLSI Networks of Spiking
  Neurons.}
\newblock {\em Neural computation}, 23(10):2457--2497, 2011.

\bibitem{Neftci2017}
Emre~O Neftci, Charles Augustine, Somnath Paul, and Georgios Detorakis.
\newblock {Event-Driven Random Back-Propagation : Enabling Neuromorphic Deep
  Learning Machines}.
\newblock {\em Froniters in Neuroscience}, 11(June):1--18, 2017.

\bibitem{Olofsson2015}
Andreas Olofsson, Tomas Nordstr{\"{o}}m, and Zain Ul-Abdin.
\newblock {Kick starting high-performance energy-efficient manycore
  architectures with Epiphany}.
\newblock {\em Conference Record - Asilomar Conference on Signals, Systems and
  Computers}, 2015-April(May):1719--1726, 2015.

\bibitem{Perez-PenaEtAl2014b}
Fernando Perez-Pena, Alejandro Linares-Barranco, and Elisabetta Chicca.
\newblock {An approach to motor control for spike-based neuromorphic robotics}.
\newblock {\em IEEE 2014 Biomedical Circuits and Systems Conference, BioCAS
  2014 - Proceedings}, pages 528--531, 2014.

\bibitem{Perez-PenaEtAl2014}
Fernando Perez-Pena, Arturo Morgado-Estevez, Teresa Serrano-Gotarredona,
  F.~Gomez-Rodriguez, V.~Ferrer-Garcia, A.~Jimenez-Fernandez, and
  A.~Linares-Barranco.
\newblock {Spike-based VITE control with dynamic vision sensor applied to an
  arm robot}.
\newblock {\em Proceedings - IEEE International Symposium on Circuits and
  Systems}, pages 463--466, 2014.

\bibitem{Pouget2000}
A~Pouget and L~H Snyder.
\newblock {Computational approaches to sensorimotor transformations.}
\newblock {\em Nature neuroscience}, 3 Suppl(November):1192--8, 2000.

\bibitem{Qiao2015}
Ning Qiao, Hesham Mostafa, Federico Corradi, Marc Osswald, Dora Sumislawska,
  Giacomo Indiveri, and Giacomo Indiveri.
\newblock {A Re-configurable On-line Learning Spiking Neuromorphic Processor
  comprising 256 neurons and 128K synapses}.
\newblock {\em Frontiers in neuroscience}, 9(February), 2015.

\bibitem{RichterEtAl2014}
M~Richter, J~Lins, S~Schneegans, Y~Sandamirskaya, and G~Sch{\"{o}}ner.
\newblock {Autonomous Neural Dynamics to Test Hypotheses in a Model of Spatial
  Language}.
\newblock In {\em The Annual Meeting of the Cognitive Science Society, CogSci},
  2014.

\bibitem{SandamirskayaConradt2013_ICANN}
Y~Sandamirskaya and J~Conradt.
\newblock {Learning Sensorimotor Transformations with Dynamic Neural Fields}.
\newblock In {\em International Conference on Artificial Neural Networks
  (ICANN)}, 2013.

\bibitem{Sandamirskaya2014}
Yulia Sandamirskaya.
\newblock {Dynamic Neural Fields as a Step Towards Cognitive Neuromorphic
  Architectures}.
\newblock {\em Frontiers in Neuroscience}, 7:276, 2013.

\bibitem{SchemmelEtAl2010}
Johannes Schemmel, Daniel Br{\"{u}}derle, Andreas Gr{\"{u}}bl, Matthias Hock,
  Karlheinz Meier, and Sebastian Millner.
\newblock {A wafer-scale neuromorphic hardware system for large-scale neural
  modeling}.
\newblock In {\em ISCAS 2010 - 2010 IEEE International Symposium on Circuits
  and Systems: Nano-Bio Circuit Fabrics and Systems}, pages 1947--1950, 2010.

\bibitem{Yin2006}
Henry~H Yin and Barbara~J Knowlton.
\newblock {The role of the basal ganglia in habit formation.}
\newblock {\em Nature reviews. Neuroscience}, 7(6):464--476, jun 2006.

\end{thebibliography}

\end{document}